\documentclass[12pt]{article}
\usepackage{hyperref}
\usepackage{amsmath}
\usepackage{graphicx}
\usepackage{amssymb} 
\newcommand{\Mat}[1]{{{\boldsymbol{#1}}}}

\def\be{\begin{equation}}
\def\ee{\end{equation}}
\def\bea{\begin{eqnarray}}
\def\eea{\end{eqnarray}}
\def\dd{\mathrm{d}}

\date{}

\title{{\bf Basic quantum mechanics for three Dirac equations in a curved spacetime}}

\author{
Mayeul Arminjon\,$^{1,2}$ and Frank Reifler\,$^3$\\
$^1$ \small\it CNRS (Section of Theoretical Physics)\\
$^2$ \small\it Laboratory ``Soils, Solids, Structures, Risks'' (CNRS \& Universit\'es de Grenoble),\\
\small\it BP 53, F-38041 Grenoble cedex 9, France.\\
\small\it $^3$ Lockheed Martin Corporation, MS2 137-205,\\ 
\small\it 199 Borton Landing Road, Moorestown, New Jersey 08057, USA.
} 

\begin{document}
\maketitle
\begin{abstract}
We study the basic quantum mechanics for a fully general set of Dirac matrices in a curved spacetime by extending Pauli's method. We further extend this study to three versions of the Dirac equation:  the standard (Dirac-Fock-Weyl or DFW) equation, and two alternative versions, both of which are based on the recently proposed linear tensor representations of the Dirac field (TRD).  We begin with the current conservation:  we show that the latter applies to any solution of the Dirac equation, iff the field of Dirac matrices $\gamma ^\mu $  satisfies a specific PDE.  This equation is always satisfied for DFW with its restricted choice for the $\gamma ^\mu $ matrices.  It similarly restricts the choice of the $\gamma ^\mu $ matrices for TRD. However, this restriction can be achieved.  The frame dependence of a general Hamiltonian operator is studied.  We show that in any given reference frame with minor restrictions on the spacetime metric, the axioms of quantum mechanics impose a unique form for the Hilbert space scalar product.  Finally, the condition for the general Dirac Hamiltonian operator to be Hermitian is derived in a general curved spacetime. For DFW, the validity of this hermiticity condition depends on the choice of the $\gamma ^\mu $ matrices.  
\\



\end{abstract}
\section{Introduction}\label{Intro}
\subsection{Aim of this work}

Quantum-mechanical effects of the classical gravitational field are currently the clearest consequence of the interplay between gravitation and the quantum. The effects which have been observed \cite{COW1975,KasevichChu1991,Nesvizhevsky2002,RiehleBorde1991,WernerStaudenmannColella1979} had been previously predicted by using the non-relativistic Schr\"odinger equation in the Newtonian gravity potential \cite{LuschikovFrank1978,OverhauserColella1974,WernerStaudenmannColella1979}, and are still described by this same non-relativistic approximation \cite{Nesvizhevsky2002,RiehleBorde1991,WestphalAbele2006}. However, one expects that the precision will increase so that, in the future, the corrections brought by the Dirac equation in a curved spacetime  (see {\it e.g.} Refs. \cite{ Boulanger-Spindel2006,HehlNi1990, Mashhoon-et-al1998, deOliveiraTiomno1962,VarjuRyder2000,A38}) should become detectable. The standard extension of the Dirac equation to a curved spacetime is due to Fock and to Weyl, and will be referred to as the ``Dirac-Fock-Weyl" (DFW) equation. In addition to this, two alternative versions of the Dirac equation in a curved spacetime were tentatively proposed in a recent work \cite{A39}. While waiting for an increase in the experimental accuracy, one would like to determine the predictions of the Dirac equation for quantum mechanics in a curved spacetime, and to check if the different versions of it might be experimentally distinguishable. \\ 

A basic feature of the Dirac equation is that its coefficients, the Dirac matrices $\gamma ^\mu$, have to satisfy the anticommmutation relation corresponding to the Lorentzian metric $g_{\mu \nu }$ on the spacetime V:
\be \label{Clifford}
\gamma ^\mu \gamma ^\nu + \gamma ^\nu \gamma ^\mu = 2g^{\mu \nu}\,{\bf 1}_4, \quad \mu ,\nu \in \{0,...,3\} \quad ({\bf 1}_4\equiv \mathrm{diag}(1,1,1,1)).
\ee
[Here $(g^{\mu \nu})$ is the inverse matrix of $(g_{\mu \nu})$.] Therefore, in a curved spacetime, or already in a flat spacetime in general coordinates, the Dirac matrices $\gamma ^\mu(X) $ depend on the spacetime point $X$, as does the metric. It follows that there is a continuum of possible choices for the field $\gamma ^\mu(X) $, all satisfying the fundamental anticommutation relation (\ref{Clifford}). The point-dependence of the metric also implies, as we will show, that there is an infinity of a priori equally valid candidates for the Hilbert space scalar product. \\

Together with the standard equation or Dirac-Fock-Weyl (DFW) equation, the two alternative equations \cite{A39} provide us with three versions of the Dirac equation in a curved spacetime, which are a priori inequivalent. The aim of the present work is to study the basic quantum mechanics: current conservation, Hilbert space inner product, and hermiticity of the Hamiltonian, for these three gravitational Dirac equations. To reach this goal, in particular to study the influence of the possible choices for the field of Dirac matrices $\gamma ^\mu (X) $, it is necessary to be able to use any possible choice of the latter. This is achieved by using the ``hermitizing matrix" $A$ of Bargmann \cite{Bargmann1932} and Pauli \cite{Pauli1933,Pauli1936}, which allows one to define the current and the scalar product for a generic (ordered) set $(\gamma ^\mu )$ of Dirac matrices \cite{A40}. To our knowledge, until now, the possible choices for the fields $\gamma ^\mu (X)$ and the Hilbert space scalar product have not been systematically investigated, {\it even for the standard (DFW) equation.} [For the latter, the different fields $\gamma ^\mu (X)$ arise due to different choices of the orthonormal tetrad field in Eq. (\ref{flat-deformed}) below.] However, it is not a priori obvious that these choices have no effect. We will show that in fact they have an essential effect. First, we will show that the axioms of quantum mechanics impose a unique form for the scalar product. The hermiticity condition has not been investigated in a general setting, again {\it even for the standard (DFW) equation.} The recent work of Leclerc \cite{Leclerc2006} studies the hermiticity of the Hamiltonian without explicitly stating restrictions on the coordinate system, but it restricts consideration to special kinds of tetrad fields, and the validity of its result is not general: we will show that, for DFW, the validity of the hermiticity condition depends on the choice of the fields $\gamma ^\mu (X)$. This indicates the presence of a uniqueness problem.\\

For example, this uniqueness problem arises for the DFW equation in a Rindler spacetime, which is a Minkowski spacetime using Rindler coordinates \cite{Oriti2000}.  In a Rindler spacetime, it is usual to choose a tetrad that is independent of the time coordinate. However, if time dependent tetrads are allowed, the time evolution of the quantum mechanical states can fail to be unitary.  How does one exclude the use of the Minkowski tetrad which is time dependent in a Rindler spacetime?  Furthermore, there is no known method for choosing this tetrad that can be generalized to general spacetimes, such that the time evolution will be unitary.  The approach taken in this paper is complementary to using tetrads, and questions such as the existence and uniqueness of a Hermitian Hamiltonian operator for a given Dirac equation---which is essential for energy eigenfunction expansions in both time-independent and time-dependent quantum mechanics---are resolved more directly by considering admissible local similarity transformations. (See Section \ref{Hamiltonian}.)  Clearly, the present investigation of the first quantized Dirac theory in a curved spacetime is a preliminary step to second quantization in a curved spacetime.  \\ 

\subsection{Status of the two alternative Dirac equations}

The two alternative versions of the Dirac equation in a curved spacetime were got by using directly the classical-quantum correspondence \cite{A39}. One version obeys \cite{A37,A39} the equivalence principle in the sense which is standard \cite{Will1993} in this context. Namely, it automatically coincides with the flat-spacetime Dirac equation ``in a local freely-falling frame," {\it i.e.,} in a coordinate system in which, at the point $X$ considered, the metric tensor reduces to the standard form $\eta_{\mu \nu } $ [with matrix $(\eta_{\mu \nu }) \equiv \mathrm{diag}(1,-1,-1,-1)$], and the {\it metric} connection vanishes. In contrast, for the DFW equation, the equivalence principle can only be established for an anholonomic frame and a spin connection \cite{A39}. The other alternative gravitational Dirac equation has a preferred reference frame, although it can be rewritten in a generally-covariant form \cite{A39}. Both of these alternative versions require the {\it tensor representation of the Dirac} (TRD) field, {\it i.e.,} the Dirac theory with a four-vector wave function and with the set of the components of the $\gamma ^\mu $ matrices building a third-order tensor \cite{A37}. In the TRD theory, the Dirac equation has the same linear form and a similar Lagrangian
as the DFW equation, \footnote{\ 
See Eq. (\ref{Dirac-general}) below for the common form of the Dirac equation and see Eq. (21) in Ref. \cite{A43} for the Lagrangian.
}
potentially giving rise to both first and second quantized theories. \\

In a Minkowski spacetime in both Cartesian and affine coordinates, the TRD theory with constant Dirac matrices has been proved \cite{A40} to be quantum-mechanically fully equivalent to the genuine Dirac theory. See also Subsect. 1.1 in Ref. \cite{A39} for a summary of the argument. Like the solutions of the genuine Dirac equation, the single particle TRD solutions have only two spin polarizations (up and down) which makes them spin-half wave functions. Moreover, just as with the genuine Dirac equation, it is straightforward to extend the single particle theory to a canonical second quantized fermion theory (see Subsect. 4.3 of Ref. \cite{Ryder1996}), with the fermion field operator built on the normalized single particle and antiparticle TRD solutions.\\

A global field of Dirac matrices $\gamma^\mu$ satisfying Eq. (\ref{Clifford}) exists (for DFW and for TRD as well) if and only if the spacetime admits spinor structure \cite{PenroseRindler1986}. It thus suggests itself to extend the investigation to curved spacetimes that admit spinor structure.  We will first present the three Dirac equations in a common framework and show that appropriate hermitizing matrices always exist (Sect. \ref{3-Dirac}). Then we shall discuss the definition of the current, the condition for its conservation, and the ways to fulfil this condition (Sect. \ref{Current}). Finally, we shall write the Hamiltonian and emphasize its frame-dependence, define the scalar product, and characterize the hermiticity condition for the Hamiltonian (Sect. \ref{Hamiltonian}). In conclusion we summarize the theorems (proved in this paper) which are common to all three Dirac equations (Sect. \ref{Conclusion}).\\

\section{Three Dirac equations in a curved spacetime}\label{3-Dirac}

\subsection{The Dirac equation with three different connections}
The three gravitational Dirac equations discussed in the \hyperref[Intro]{Introduction} have a common form:
\be\label{Dirac-general}
\gamma ^\mu D_\mu\psi=-im\psi,\\
\ee

\vspace{3mm}
\noindent where $\gamma ^\mu =\gamma ^\mu (X)$ ($\mu =0,...,3$) is a field of $4\times 4$ complex matrices defined on the spacetime manifold V, satisfying the anticommmutation relation (\ref{Clifford});
and where $i=\sqrt{-1}$, $m$ is the rest-mass of the particle (setting $\hbar=1=c$: otherwise, replace $m$ by $mc/\hbar$), $\psi$ is either a quadruplet of four scalar fields (DFW) or a four-vector field (TRD), and $D_\mu $ is a covariant derivative, associated with a specific {\it connection}.\\

For the two alternative equations based on the tensor representation of the Dirac field (TRD), this is an affine connection ({\it i.e.,} a connection associated with the tangent bundle \cite{ChernChenLam1999, DieudonneTome3}):
\be\label{D_mu psi^nu}
(D_\mu \psi)^\nu \equiv \partial_\mu \psi^\nu  +\Delta ^\nu _{\rho \mu }\psi ^\rho.
\ee
In the TRD equations, the affine connection is extended to the complexified tangent bundle. More precisely, for one of the two TRD equations, henceforth denoted TRD-1, this is simply the Levi-Civita connection. That is, the $\Delta ^\nu _{\rho \mu }$'s are the (second-kind) Christoffel symbols associated with the spacetime metric $g_{\mu \nu }$:
\be\label{Levi-Civita}
\Delta ^\nu _{\rho \mu } \equiv  \left\{^\nu _{\rho \mu } \right\} .
\ee
This is the one which obeys the equivalence principle, in the sense stated in the \hyperref[Intro]{Introduction}. For the other TRD equation (TRD-2), the connection $\Delta $ is defined from the {\it spatial} Levi-Civita connection in an assumed preferred reference frame (a congruence of observers \cite{Cattaneo}, here endowed with a preferred time coordinate \cite{A39,A45}, denoted E. In any coordinates adapted to E, we have
\be\label{Delta-connection-etherframe}
\Delta^{\nu} _{\rho \mu} \equiv \left\{ \begin{array}{ll} 0 &  \mathrm{if}\ \nu=0 \mathrm{\ or\ }\mu =0 \mathrm{\ or\ }\rho =0\\
& \\
 G ^{j} _{l k} & \mathrm{if\ }\nu=j \ \mathrm{and\ }\mu=k\ \mathrm{and\ }\rho = l \in \{1,2,3\},
\end{array} \right.
\ee
the $G ^j _{lk}$'s ($j,k,l \in \{1,2,3\}$) being the Christoffel symbols of the spatial metric $\Mat{h}$ \cite{Cattaneo,L&L,Moeller} in that preferred frame E. In a general coordinate system, the $\Delta ^\nu _{\rho \mu }$'s are obtained from the coefficients (\ref{Delta-connection-etherframe}), by using the transformation law of affine connections \cite{Doubrovine1982}: it is proved in Ref. \cite{A39} that this does define a unique connection (which is torsionless). Equation TRD-2, being based on that ``preferred-frame connection," has clearly a more speculative character from the physical point of view. Note that, in the case of a {\it static} (and non-flat) spacetime, we do have one preferred reference frame with a preferred time coordinate \cite{A39}. Also note that, in the case of a {\it flat} spacetime, special relativity does apply. Thus, in that case, we must find that we may take any inertial frame as ``the preferred frame E." And indeed, if we take for E any inertial frame, the connection defined by (\ref{Delta-connection-etherframe}) coincides with the Levi-Civita connection associated with the flat metric (with, in particular, all $\Delta^{\nu} _{\rho \mu}$ 's being zero if the coordinates are Cartesian). In other words, the two equations TRD-1 and TRD-2 coincide in the case of a flat spacetime.\\

For the DFW equation, the connection is the ``spin connection" acting on the trivial bundle $\mathrm{V} \times {\sf C^4}$.
\footnote{\
It is well known that a given spacetime V need not admit a spinor structure. It was proved by Geroch that a four-dimensional noncompact spacetime admits a spinor structure if and only if it admits a global tetrad field \cite{Geroch1968}. In that case, both spinor bundles and tangent bundles are trivial \cite{Geroch1968,Isham1978}. Penrose and Rindler argue that these are the only spacetimes of interest \cite{PenroseRindler1986}. Then it is easy to show that one may define a global field of Dirac matrices $\gamma^\mu$  satisfying Eq. (\ref{Clifford}) above, for DFW and for TRD as well.
}
It is built from the ``spin matrices" $\Gamma _\mu $ \cite{BrillWheeler1957+Corr}: we shall take the positive sign,
\be\label{D_mu psi}
D_\mu \psi  \equiv \partial _\mu \psi + \Gamma _\mu \psi,
\ee
so that (\ref{D_mu psi^nu}) takes the form (\ref{D_mu psi}) if we associate matrices $\Gamma  _\mu $ to connection coefficients (or conversely) by
\be\label{Gamma_mu-affine}
\left(\Gamma  _\mu\right)^\nu _{\ \, \rho } \equiv \Delta^{\nu} _{\rho \mu}.
\ee
However, except in the Majorana representation where the $\Gamma _\mu $'s are real, the spin connection matrices $\Gamma _\mu $ of the DFW equation are generally complex, and depend on the set of fields $(\gamma ^\mu) \ (\mu =0,...,3)$.\\

The transformation of the Dirac equation (\ref{Dirac-general}) for a coordinate change depends also on the version: for the two TRD equations \cite{A39}, the wave function $\psi $ is a {\it spacetime vector} (or four-vector), hence it transforms thus:
\be \label{psi-vector}
\psi '=L\psi \quad (\psi '^\mu = L^\mu_{\ \nu}\, \psi ^\nu), \qquad L^\mu_{\ \nu} \equiv \frac{\partial x'^\mu }{\partial x^\nu },
\ee
and the threefold array of the components of the Dirac matrices, $\gamma^{\mu \rho} _\nu\equiv \left(\gamma^\mu \right)^\rho _{\ \nu}$, builds a $\ {\it (^2 _1)}$ {\it tensor,} thus
\be \label{gamma-(^2_1)tensor}
\gamma '^\mu = L^\mu_{\ \sigma }\, L \gamma ^\sigma  L^{-1}, \qquad \mathrm{or}\quad \gamma'^{\mu \rho} _\nu = L^\mu_{\ \sigma }\,L^\rho _{\ \tau }\,\left(L^{-1}\right)^\chi_{\ \nu  }\gamma^{\sigma \tau } _\chi .
\ee
In contrast, for the DFW equation \cite{BrillWheeler1957+Corr}, the wave function transforms like a {\it scalar:}
\be\label{psi-scalar} 
\psi'^{\mu} ((x'^{\nu})) = \psi ^{\mu} ((x ^{\nu})),
\ee
and the set of the $\gamma ^\mu$ matrices transforms like a {\it four-vector:}
\be\label{gamma-vector}
\gamma'^\mu =L^\mu_{\ \nu}\, \gamma^\nu.
\ee
Either of these two transformation modes leaves the Dirac equation (\ref{Dirac-general}) form invariant after a general coordinate change:  for the TRD equations, each side of (\ref{Dirac-general}) is a four-vector; whereas, for DFW, it is an object with four scalar components. 

\subsection{A common tool: the hermitizing matrices}

To define the current, we shall use the hermitizing matrix for a general set $(\gamma  ^\mu )$ of Dirac matrices. This is a nonzero $4 \times 4$ complex matrix $A$ such that\be\label{hermitizing-A}
A^\dagger = A, \qquad (A\gamma ^\mu )^\dagger = A\gamma ^\mu \quad \mu =0, ...,3,
\ee
where $M^\dagger\equiv M^{*\,T}$ denotes the Hermitian conjugate of a matrix $M$. Moreover, we shall see that the Hamiltonian operator associated with (\ref{Dirac-general}) depends on the field of matrices $\alpha ^\mu $, with
\be \label{alpha}
\alpha ^0 \equiv \gamma ^0/g^{00}, \qquad \alpha ^j \equiv \gamma ^0\gamma ^j/g^{00}.
\ee
Therefore, to define a relevant scalar product, we will also need a hermitizing matrix, denoted $B$, for the set $(\alpha ^\mu )$:
\be\label{hermitizing-B}
B^\dagger = B, \qquad (B\alpha  ^\mu )^\dagger = B\alpha  ^\mu \quad \mu =0, ...,3,
\ee
and we will need that it be {\it positive-definite}. The existence of $A$ and $B$ is ensured by the following result \cite{A40}:
\footnote{\ 
The existence of a hermitizing matrix $A$ had been already proved by Pauli \cite{Pauli1933,Pauli1936}, though in a less general case and with less complete results (see Ref. \cite{A40}).
}
\\

\noindent {\bf Theorem \cite{A40}-6.}  {\it Fix any point $X$ in the spacetime.  For any set of matrices $\gamma ^\mu$  satisfying the general anticommutation formula (\ref{Clifford}), there exists a hermitizing matrix $A$  for the matrices $\gamma ^\mu$.  The matrix  $A$ is nonsingular and unique, up to a real scale factor. Similarly, a nonsingular hermitizing matrix $B\equiv A\gamma ^0$ for the $\alpha ^\mu$'s exists and is unique, up to a real scale factor. If, furthermore, the coordinate system is an admissible one, {\it i.e.,} if $g_{00}>0$ and the $3\times 3$ matrix $(g_{jk})\ (j,k=1,2,3)$ is negative definite, then $B\equiv A\gamma ^0$  is either a positive or negative definite matrix.  The sign of the matrix $A$ can be chosen such that $B\equiv A\gamma ^0$ is a positive definite matrix. }\\

The proof of this theorem in Ref. \cite{A40} is directly valid for TRD, because that proof uses the tensor transformation of the Dirac matrices, Eq. (\ref{gamma-(^2_1)tensor}) here, but this is not in an essential way. For DFW, the field of the matrices $\gamma ^\mu $ satisfying the anticommutation relation (\ref{Clifford}) is defined from an orthonormal tetrad $(u_\alpha )$, with $u_\alpha   \equiv  a^\mu  _{\ \,\alpha} \, \frac{\partial }{\partial x^\mu  } \quad (\alpha =0,...,3)$, by \cite{BrillWheeler1957+Corr,Leclerc2006,deOliveiraTiomno1962}
\be \label{flat-deformed}
  \gamma ^\mu = a^\mu_{\ \,\alpha}  \ \gamma ^{ \natural \alpha} ,
\ee
where $(\gamma ^{ \natural \alpha} )$ could be any set of constant Dirac matrices, {\it i.e.,} obeying Eq. (\ref{Clifford}) with $(g^{\mu \nu })=(\eta _{\mu \nu })^{-1}=(\eta _{\mu \nu })\equiv \mathrm{diag}(1,-1,-1,-1)$. It is immediate to check that, due to the fact that the matrix $a\equiv (a^\mu_{\ \,\alpha})$ and its inverse $b\equiv (b^\alpha _{\ \,\mu })$ are real, any matrix $A^\natural$ that is hermitizing for the $\gamma ^{ \natural \alpha} $'s is also hermitizing for the $\gamma ^\mu $'s, and conversely. Using this fact, it is straightforward to modify the proof of Theorem 6 in Ref. \cite{A40} so that it applies to DFW.

\vspace{5mm}
Thus, at each point $X$ in the spacetime V, for both TRD and DFW theories, we have hermitizing matrices $A(X)$ and $B(X)$ for the sets $(\gamma ^\mu (X))$ and $(\alpha  ^\mu(X))$, respectively. For TRD, the components $A_{\mu \nu }$ of matrix $A$ build a {\it covariant second-order tensor} \cite{A40}. For DFW, we may choose $A=A^\natural$. However, since $A$ is uniquely defined only up to a real factor, we might also multiply by a positive real scalar field $\lambda $:
\be\label{A=lambda A^natural}
A=\lambda (X) A^\natural.
\ee
Because the ``flat" matrices $\gamma ^{\natural \alpha }$ of Eq. (\ref{flat-deformed}), hence also $A^\natural$, are constant matrices in the DFW theory, it follows from (\ref{A=lambda A^natural}) for each point $X \in \mathrm{V}$ that $A(X)$ is a matrix with scalar components in the DFW theory. At each spacetime point, the matrix
\be\label{B=A gamma0}
B\equiv A\gamma ^0
\ee
is positive definite for both the TRD and DFW theories by the Theorem above. Note that, by (\ref{Clifford}) and (\ref{alpha}), we get conversely
\be\label{A=B alpha0} 
A=B\alpha ^0.
\ee 

\vspace{4mm}
Using Eq. (\ref{hermitizing-A}), we define a Hermitian product between complex four-vectors (or quadruplets of complex scalars) $u,v $ by setting \be\label{A-product}
(u ,v ) \equiv A_{\rho \nu } u ^{\rho*} v^\nu = u^\dagger A v,
\ee
and in the same way, we define a positive-definite Hermitian product, by setting 
\be\label{B-product}
(u : v ) \equiv B_{\rho \nu } u ^{\rho*} v^\nu = u^\dagger B v.
\ee
In the TRD theory, the four-vectors $u,v \in \mathrm{T}_{\sf C}\mathrm{V}_X$, where $\mathrm{T}_{\sf C}\mathrm{V}_X$ is the complexification of the tangent space $\mathrm{T}\mathrm{V}_X$ at $X \in \mathrm{V}$. 
\footnote{\
In the TRD theory, Eqs. (\ref{A-product})$_2$ and (\ref{B-product})$_2$ exploit an abuse of notation that views $A$ and $B$ in two different ways: first as tensors and then as matrices representing Hermitian forms \cite{A40}. Compare these equations with the similar equation using the spacetime metric $G$ viewed first as a tensor and then as a matrix representing a quadratic form. In this case we would have $g_{\rho \nu }u^\rho v^\nu =u^T Gv$  for  $u,v \in \mathrm{T}\mathrm{V}_X$.  This is harmless as long as the covariance is being carefully checked. 
}
It results from (\ref{hermitizing-A}) [resp. from (\ref{hermitizing-B})] that each of the $\gamma ^\mu$ [resp. $\alpha  ^\mu $] matrices is a Hermitian operator for the product (\ref{A-product}) [resp. (\ref{B-product})], that is,
\be\label{gamma-hermitian-A-product}
(\gamma ^\mu u,v) =(u,\gamma ^\mu v), \quad \mu =0, ...,3,
\ee
\be\label{alpha-hermitian-B-product}
(\alpha  ^\mu u: v) =(u: \alpha  ^\mu v), \quad \mu =0, ...,3.
\ee
Finally, we note that, due to Eqs. (\ref{B=A gamma0}) and (\ref{A=B alpha0}), we have
\be \label{,to:}
(u,v) = (\alpha ^0 u:v)=(u:\alpha ^0 v)
\ee
and
\be \label{:to,}
(u:v) = (\gamma  ^0 u,v)=(u,\gamma  ^0 v).
\ee

\section{Current conservation}\label{Current}

\subsection{Definition of the current}

Obviously, the definition of the current should involve the Dirac matrices. There are an infinity of sets of fields $(\gamma ^\mu) $ that satisfy the anticommutation relation (\ref{Clifford}) in the given curved spacetime $(\mathrm{V},g_{\mu \nu })$. In the case of a flat spacetime, it is natural to assume that the matrices $\gamma ^\mu $ are {\it constant} in Cartesian coordinates (such that the metric has the standard form, $g_{\mu \nu }=\eta _{\mu \nu }$). 
\footnote{\
We note in passing that, for TRD, this is equivalent to say that the $\gamma ^\mu $'s are covariantly constant: $D_\sigma  \gamma ^\mu =0$, or explicitly $D_\sigma  \gamma ^{\mu \rho }_\nu =0$. Whereas, for DFW, the derivatives $D_\sigma  \gamma ^\mu$ are {\it always} zero \cite{BrillWheeler1957+Corr}: contrary to the Levi-Civita connection, the spin connection does not recognize anything special in the case of constant Dirac matrices in the flat situation. 
}
In the latter case, the current is unambiguously defined as
\be\label{J-mu-standard}
J^\mu \equiv (\gamma ^\mu \psi ,\psi ) = A_{\rho \nu } \left(\gamma^{\mu\, *} \right)^{\rho}  _{\ \sigma}\psi ^{\sigma *} \psi ^{\nu },
\ee
or equivalently [using (\ref{hermitizing-A})]:
\be \label{J-mu-standard-matrix}
J^\mu =  \psi ^\dagger \gamma ^{\mu\,\dagger}\,A \,\psi = \psi ^\dagger B^\mu \psi,\qquad (B^\mu \equiv A\gamma ^\mu).
\ee
Indeed, this definition coincides with the usual one \cite{BjorkenDrell1964,Schulten1999} for the standard set $(\gamma^{\sharp \mu })$ of ``flat" Dirac matrices---for which set $A\equiv \gamma^{\sharp 0}$ turns out to be a hermitizing matrix. The case with the standard set has been thoroughly investigated in the literature, in particular the current (\ref{J-mu-standard-matrix}) [with $A\equiv \gamma^{\sharp 0}$] is then derived from the Dirac Lagrangian, so that there is no ambiguity. In addition, it turns out \cite{A40} that the current (\ref{J-mu-standard-matrix}) is actually {\it independent on the choice of the Dirac matrices:} if one changes one set $(\gamma ^\mu)$ for another one $(\tilde {\gamma }^\mu)$ [satisfying the {\it same} anticommutation relation (\ref{Clifford}) as does $(\gamma ^\mu)$], then the second set can be obtained from the first one by a {\it similarity transformation:}
\be \label{similarity-gamma}
\exists S \in {\sf GL}(4,{\sf C}):\qquad \tilde{\gamma} ^\mu =  S\gamma ^\mu S^{-1}, \quad \mu =0,...,3,
\ee
for which the solutions of the flat-spacetime Dirac equation exchange by
\be \label{similarity-psi-S}
\tilde{\psi }=  S\psi.
\ee
The hermitizing matrix is transformed thus:
\be \label{similarity-A}
\tilde {A}= (S^{-1})^\dagger A S^{-1}=(S^\dagger)^{-1} A S^{-1},
\ee
and this leads indeed to the invariant relation
\be \label{J-tilde}
\tilde {J}^\mu \equiv \tilde {\psi} ^\dagger \tilde {\gamma} ^{\mu\,\dagger}\,\tilde {A} \,\tilde {\psi}= J^\mu.\\
\ee
The definition (\ref{J-mu-standard-matrix}) of the current is thus the right one in the flat case, and it is generally-covariant, the current being indeed a {\it four-vector,} for TRD and for DFW as well---as it results immediately from the transformation behaviours of its ingredients $\gamma ^\mu ,\psi $, and $A$ in the two theories. Therefore, we assume (\ref{J-mu-standard-matrix}) [or (\ref{J-mu-standard})] as the definition of the current in the general case of a curved spacetime. We note that the invariance (\ref{J-tilde}) of the current after a similarity (\ref{similarity-gamma})-(\ref{similarity-A}) remains in force in that general case, even if matrix $S$ depends on the event $X$. We check from (\ref{J-mu-standard-matrix}), (\ref{B=A gamma0}) and  the positive definiteness of matrix $B$ that the probability density is $J^0 =\psi^\dagger B\psi  \geq 0$, with $J^0 >0$ if $\psi \ne 0$, as it must be.

\subsection{Characteristic condition for current conservation}\label{conserv-condition}

From the current definition (\ref{J-mu-standard-matrix}) and Leibniz' rule, one gets trivially
\be \label{Dmutrivial}
D_\mu J^\mu=(D_\mu \psi) ^\dagger B^\mu \psi +\psi ^\dagger (D_\mu B^\mu) \psi + \psi ^\dagger  B^\mu D_\mu\psi,
\ee
provided the derivative $D_\mu B^\mu$ is well defined. Note that, for TRD, $B^\mu _{\nu \rho }\equiv (A\gamma ^\mu )_{\nu \rho }=A_{\nu \sigma  }\gamma ^{\mu \sigma }_\rho $ is a spacetime tensor. 
\footnote{\
More exactly, the first contravariant index $\mu $ in $\gamma ^{\mu \sigma }_\rho$ corresponds to tensor components of the (real) tangent space, as indicated by the (real) anticommutation relation (\ref{Clifford}). In contrast, the second contravariant index, $\sigma $, as well as the covariant index, $\rho $, correspond to tensor components of the complex tangent space, as apparent from Eq. (\ref{J-mu-standard}). In other words, the tensor $\gamma ^{\mu \sigma }_\rho$ at each spacetime point $X$ is an element of the tensor space $ \mathrm{T}\mathrm{V}_X	\otimes \mathrm{T}_{\sf C}\mathrm{V}_X \otimes \mathrm{T}_{\sf C}^\circ \mathrm{V}_X$, where $\mathrm{T}_{\sf C}\mathrm{V}_X$  and $\mathrm{T}_{\sf C}^\circ \mathrm{V}_X$ are the complexifications of the real tangent space $ \mathrm{T}\mathrm{V}_X$  and its dual $\mathrm{T}^\circ \mathrm{V}_X$. Similarly, from (\ref{A-product}), the tensor $A_{\nu \sigma  }$ belongs to $\mathrm{T}_{\sf C}^\circ\mathrm{V}_X \otimes \mathrm{T}_{\sf C}^\circ\mathrm{V}_X$, so that the contracted product $B^\mu _{\nu \rho }=A_{\nu \sigma  }\gamma ^{\mu \sigma }_\rho $ makes sense and belongs to $ \mathrm{T}\mathrm{V}_X	\otimes \mathrm{T}_{\sf C}^\circ\mathrm{V}_X \otimes \mathrm{T}_{\sf C}^\circ\mathrm{V}_X$. Now, for different types of indices, we may use different connections. For ``complex" indices, we are using the connection $\Delta ^\nu _{\rho \mu } $, Eq. (\ref{D_mu psi^nu}), which differs from the metric connection $  \left\{^\nu _{\rho \mu } \right\}$ for TRD-2; but, {\it for ``real" indices, we shall use the metric connection for both TRD-1 and TRD-2.} Leibniz' rule still applies.

}
$D_\mu B^\mu$ is then the matrix with components $D_\mu B^\mu _{\nu \rho }$, the latter being defined in the standard way from the relevant affine connections, thus
\be\label{D_mu B^mu}
D_\mu B^\mu _{\nu \rho }=\partial_\mu B^\mu _{\nu \rho } + \left\{^\mu _{\sigma  \mu } \right\}B^\sigma  _{\nu \rho } - \Delta ^\sigma  _{\nu \mu }B^\mu _{\sigma \rho } -\Delta ^\sigma  _{\rho \mu }B^\mu _{\nu \sigma }.
\ee
For DFW, $\gamma ^\nu $ is a ``spinor vector" [Eq. (\ref{gamma-vector})], whose spin covariant derivative is defined as \cite{BrillWheeler1957+Corr,PenroseRindler1986}
\be\label{D_mu gamma^nu}
D_\mu \gamma ^\nu \equiv \partial _\mu \gamma ^\nu + \left\{^\nu _{\rho \mu } \right\}\gamma ^\rho + \left[\Gamma _\mu,\gamma ^\nu \right]
\ee
(where $[M,N]\equiv MN-NM$), and this is known to be {\it zero:}
\be\label{D_mu gamma^nu=0}
D_\mu \gamma ^\nu =0 \qquad (\mathrm{DFW});
\ee
while the derivative of a ``spinor scalar" like the hermitizing matrix $A$ is defined to be
\be\label{D_mu A}
D_\mu A \equiv \partial _\mu A - A\Gamma _\mu - \Gamma _\mu^\dagger A,
\ee
ensuring that $D_\mu (A^\dagger )=(D_\mu A)^\dagger$. Since the spin matrices have the form \cite{BrillWheeler1957+Corr}
\be \label{Gamma_mu}
\Gamma _\mu = c_{\lambda \nu \mu }  s^{\lambda \nu} \qquad (\mathrm{DFW})
\ee
with real coefficients $c_{\lambda \nu \mu }$, and where $ s^{\lambda \nu} \equiv \frac{1}{2} \left (\gamma ^\lambda \gamma ^\nu - \gamma ^\nu \gamma ^\lambda \right )$, it follows from (\ref{D_mu A}) and the hermitizing character (\ref{hermitizing-A}) of $A$ that
\be 
D_\mu A - \partial _\mu A =0 \qquad (\mathrm{DFW}).
\ee
Moreover, for DFW, we may choose $\lambda \equiv 1$ in Eq. (\ref{A=lambda A^natural}), {\it i.e.,} $A\equiv A^\natural$, so that we get
\be\label{D_mu A=0}
D_\mu A =0 \qquad (\mathrm{DFW},\quad A\equiv A^\natural),
\ee
and from (\ref{D_mu gamma^nu=0}), by using Leibniz' rule:
\be\label{D_mu(B_mu)=0-DFW}
D_\mu B^\mu=0\qquad (\mathrm{DFW},\quad A\equiv A^\natural).\\
\ee
Note that, if one uses also matrices $\Gamma _\mu $ for TRD, as defined from the connection coefficients [Eq. (\ref{Gamma_mu-affine})], then the definitions (\ref{D_mu gamma^nu}) and (\ref{D_mu A}), and in fact all definitions of covariant derivatives used in DFW theory \cite{PenroseRindler1986}, {\it also apply to TRD}---but, of course, not in general the results (\ref{D_mu gamma^nu=0}) and (\ref{D_mu(B_mu)=0-DFW}), which are specific to DFW theory.

\vspace{6mm}
Since $A$ and $B^\mu$ are hermitian matrices, the first term on the r.h.s. of (\ref{Dmutrivial}) is 
\be\label{term1-DmuJmu}
(D_\mu \psi) ^\dagger B^\mu \psi=(B^\mu D_\mu \psi)^\dagger \psi =[A(\gamma ^\mu D_\mu \psi)]^\dagger \psi=(\gamma ^\mu D_\mu \psi)^\dagger A\psi.
\ee
Therefore, if the (relevant) Dirac equation (\ref{Dirac-general}) is satisfied, then the two extreme terms on the r.h.s. of (\ref{Dmutrivial}) cancel one another:
\be\label{extreme-cancel}
(\gamma ^\mu D_\mu \psi)^\dagger A\psi+\psi ^\dagger A(\gamma ^\mu D_\mu \psi)=im \psi^\dagger A\psi+\psi ^\dagger A(-im\psi )=0,
\ee
whence from (\ref{Dmutrivial}):
\be \label{DmuJmu-Dirac}
D_\mu J^\mu=\psi ^\dagger (D_\mu B^\mu) \psi .
\ee

\vspace{3mm}
\noindent Our definition of the covariant derivative uses the metric connection for both TRD-1 and TRD-2 when ``real" indices are concerned (see Footnote 6)---as is the case for the index $\mu $ in the foregoing equation. That is, for TRD, as well as for DFW, the connection acting on the probability current $J^\mu $ is the Levi-Civita connection. Hence, it  follows that, for TRD as well as for DFW, the condition for conservation of the probability current is the same as in a Riemannian spacetime, namely $D_\mu J^\mu=0$. Hence we can state the following result: \\

\paragraph{Theorem 1.}\label{Theorem1} {\it Consider the general Dirac equation (\ref{Dirac-general}), thus either DFW or any of the two TRD equations. In order that any $\psi $ solution of (\ref{Dirac-general}) satisfy the current conservation
\be\label{current-conservation}
D_\mu J^\mu=0,
\ee 
it is necessary and sufficient that} 
\be\label{D_mu(B_mu)=0}
D_\mu B^\mu=0\qquad (B^\mu \equiv A\gamma ^\mu).\\
\ee

The reasoning which was used to get Theorem 1, hence also this theorem itself, extend immediately to the {\it transition probability currents}
\be \label{J-mu-transition-matrix}
K^\mu(\psi ,\psi ') \equiv  \psi ^\dagger B^\mu \psi'.\\
\ee

\vspace{5mm}
We note also that Eq. (\ref{D_mu(B_mu)=0-DFW}) gives to Theorem 1 the following\\

\paragraph{Corollary 1.}\label{Corollary1} {\it For DFW theory, the hermitizing matrix field can be imposed to be $A\equiv A^\natural $, with $A^\natural $ a constant hermitizing matrix for the ``flat" constant Dirac matrices $\gamma ^{ \natural \alpha}$ of Eq. (\ref{flat-deformed}). Then the current conservation applies to any solution of the DFW equation.}\\

The current conservation usually stated for DFW theory ({\it e.g.} \cite{BrillWheeler1957+Corr,Leclerc2006,deOliveiraTiomno1962}) applies only to the case where the ``flat" Dirac matrices $\gamma ^{ \natural \alpha}$ of Eq. (\ref{flat-deformed}) are the {\it standard choices} of Dirac matrices, which we shall denote generically $\gamma ^{ \sharp \alpha}$, and for which $\gamma ^{ \sharp 0}$ turns out to be hermitizing. \{These choices of Dirac matrices are related by similarity  transformations that are unitary \cite{Pal2007}, and not by general similarity transformations (\ref{similarity-gamma}) that leave the anticommutation formula (\ref{Clifford}) invariant.\} This is a particular case of Corollary 1.

\subsection{Modified Dirac equation with conserved current}\label{ModifiedDirac}

Due to the r.h.s. of (\ref{DmuJmu-Dirac}), the current is in general not conserved for solutions of the Dirac equation (\ref{Dirac-general}) (the {\it general} one, {\it i.e.,} DFW or TRD as well, although for DFW the natural choice $A\equiv A^\natural$ does ensure the current conservation). However, we can {\it modify} this equation to the following one:\\
\be\label{Dirac-general-modified}
\gamma ^\mu D_\mu\psi=-im\psi-\frac{1}{2}A^{-1}(D_\mu B^\mu )\psi,\\
\ee

\vspace{2mm}
\noindent so as to conserve the transition probability current (\ref{J-mu-transition-matrix}). Indeed we have the\\ 

\paragraph{Theorem 2.}\label{Theorem2} {\it For any pair $(\psi,\psi ') $ of solutions of the modified Dirac equation (\ref{Dirac-general-modified}), the transition probability current (\ref{J-mu-transition-matrix}) is conserved.}\\

{\it Proof.} Similarly to Eqs. (\ref{Dmutrivial}) and (\ref{term1-DmuJmu}), we get from the definition (\ref{J-mu-transition-matrix}):
\be \label{Dmu-K^mu}
D_\mu K^\mu=(\gamma ^\mu D_\mu \psi)^\dagger A\psi' +\psi ^\dagger (D_\mu B^\mu) \psi' + \psi ^\dagger  A\gamma ^\mu D_\mu\psi'.
\ee
Then, if the modified Dirac equation (\ref{Dirac-general-modified}) is satisfied by $\psi $ and by $\psi '$ as well, the contributions coming from $-im\psi $ and $-im\psi '$ to the two extreme terms on the r.h.s. of (\ref{Dmu-K^mu}) cancel one another, as in Eq. (\ref{extreme-cancel}). Thus we are left with
\be \label{Dmu-K^mu-Dirac-modif}
D_\mu K^\mu=(C \psi)^\dagger A\psi' +\psi ^\dagger (D_\mu B^\mu) \psi' + \psi ^\dagger  AC\psi', 
\ee
where
\be
 C \equiv -\frac{1}{2}A^{-1}(D_\mu B^\mu ).
\ee
But since $A=A^\dagger$, we have $(A^{-1})^\dagger=A^{-1}$, hence [noting that $(D_\mu B^\mu )^\dagger =D_\mu (B^{\mu \dagger})=D_\mu B^\mu$]:
\be
(C \psi)^\dagger = \psi^\dagger C^\dagger =-\frac{1}{2}\psi^\dagger (D_\mu B^\mu )A^{-1},
\ee
so that the r.h.s. of (\ref{Dmu-K^mu-Dirac-modif}) vanishes, as claimed by \hyperref[Theorem2]{Theorem 2}. Q.E.D.\\

Note that the modified Dirac equation (\ref{Dirac-general-modified}) coincides with the normal one (\ref{Dirac-general}) for all $\psi $, iff the condition for current conservation (\ref{D_mu(B_mu)=0}) is satisfied. Therefore, Eq. (\ref{Dirac-general-modified}) is really the adequate modification of (\ref{Dirac-general}) to get the current conserved in the general case. However, it is the normal Dirac equation, not the modified one, that (in the TRD case) has been derived from the classical-quantum correspondence \cite{A39}. Moreover, the TRD-1 version [based on the Levi-Civita connection (\ref{Levi-Civita})] of the normal equation obeys the equivalence principle in the precise sense of the \hyperref[Intro]{Introduction}; whereas it is not the case for the corresponding version of Eq. (\ref{Dirac-general-modified}), because the vanishing of the connection (\ref{Levi-Civita}) does not imply the validity of the condition (\ref{D_mu(B_mu)=0}). Thus, one may feel that the physically relevant equation remains the normal one (\ref{Dirac-general}). Since the current conservation is very important, this option means that {\it not all possible fields $\gamma ^\mu ,A$ are physically admissible, but merely the ones which, in addition to the anticommutation relation (\ref{Clifford}), satisfy condition (\ref{D_mu(B_mu)=0}). Such systems will be called admissible.} This, after all, is just an extension to the general case of the statement made for the flat case, that any relevant field $\gamma ^\mu $ (and hence also the field $A$) has to be constant in Cartesian coordinates: if one selects the gamma field at random, the condition (\ref{D_mu(B_mu)=0}) and the current conservation do not generally apply to the solutions of (\ref{Dirac-general}) {\it even in a flat spacetime}---except for DFW. 

\subsection{Similarity transformations under which the Dirac equation is covariant}\label{Dirac-covariant similarity}

Consider a local similarity transformation of the coefficient fields: starting with a ``fiduciary" set of fields, $\gamma^\mu,A,B^\mu \equiv A\gamma^\mu$, let us apply an event-varying similarity transformation $S(X)$ [we exchange $S$ for $S^{-1}$ w.r.t. (\ref{similarity-gamma})-(\ref{similarity-A}), for convenience]:
\be \label{similarity-gamma-invert}
\tilde{\gamma} ^\mu =   S^{-1}\gamma ^\mu S,
\ee
\be \label{similarity-A-Bmu}
\tilde{A}=  S^\dagger  A S, \qquad \tilde{B}^\mu\equiv  \tilde{A}\tilde{\gamma}^\mu =S^\dagger  B^\mu S.
\ee

\vspace{4mm}
\noindent For DFW, the similarity transformations must be restricted to the form $\mathsf{S}(L)$ with $L\mapsto \pm \mathsf{S}(L)$  the spinor representation, see Eq. (\ref{S=S(L)}). As is well known, the DFW equation is covariant under all such similarities \cite{BrillWheeler1957+Corr}. For TRD, we may ask whether the general Dirac equation---either the normal one (\ref{Dirac-general}) or the modified one (\ref{Dirac-general-modified})---is covariant under a such transformation, if one simultaneously imposes that the wave function must transform naturally as
\be \label{similarity-psi}
\tilde{\psi} =S^{-1}\psi.
\ee
We examine three questions: i) $\psi $ obeying the modified equation, when does $\tilde{\psi}$ obey the normal one? ii) $\psi $ obeying the normal equation, when does $\tilde{\psi}$ obey it also? and iii) $\psi $ obeying the modified equation, when does $\tilde{\psi}$ obey it also? The first question occurs most naturally after the discussion at the end of subsect. \ref{ModifiedDirac}. The answer to it is given by\\

\paragraph{Theorem 3.}\label{Theorem3} {\it In order that $\tilde{\psi} $ obey (\ref{Dirac-general}) each time that $\psi$ obeys (\ref{Dirac-general-modified}), it is necessary and sufficient that}
\be\label{Z=0}
Z\equiv B^\mu D_\mu S +\frac{1}{2} (D_\mu B^\mu) S=0.
\ee
{\it Moreover, if $Z=0$, the transformed fields satisfy condition (\ref{D_mu(B_mu)=0}), hence the current conservation is valid, after the transformation, with the normal Dirac equation (\ref{Dirac-general})}.\\

{\it Proof.} Entering (\ref{similarity-psi}) into (\ref{Dirac-general-modified}), we get successively
\bea
0 & = & \gamma ^\mu S (D_\mu \tilde{\psi}) +\gamma ^\mu (D_\mu S) \tilde{\psi}+\frac{1}{2}A^{-1}(D_\mu B^\mu )S\tilde{\psi}+imS\tilde{\psi}\nonumber\\
0 & = & S\left[(S^{-1}\gamma ^\mu S)D_\mu +im\right]\tilde{\psi} +\gamma ^\mu (D_\mu S) \tilde{\psi} +\frac{1}{2}A^{-1}(D_\mu B^\mu )S\tilde{\psi}\label{similarity-Dirac-modified}\\
0 & = & S \tilde{\mathcal{D}}\tilde{\psi} +A^{-1}Z \tilde{\psi},
\eea
the latter because $A^{-1}B^\mu =\gamma ^\mu $, and where $\tilde{\mathcal{D}}$ is the normal Dirac operator with the transformed fields $\tilde{\gamma} ^\mu$, the normal Dirac operator being $\mathcal{D}\equiv \gamma ^\mu D_\mu+im$. This proves the first part of Theorem 3. It remains to check that
\be\label{D_mu(tildeB_mu)=0}
Z=0 \quad \Rightarrow \quad D_\mu \tilde {B}^\mu=0.
\ee
Entering into (\ref{D_mu(tildeB_mu)=0})$_2$ the definition (\ref{similarity-A-Bmu}) of $\tilde {B}^\mu$ as a transformed quantity, yields
\be
S^\dagger  B^\mu (D_\mu S)+ (D_\mu S^\dagger )B^\mu S=-S^\dagger  (D_\mu B^\mu)  S.
\ee
Thus, (\ref{D_mu(tildeB_mu)=0})$_2$ is equivalent to
\be\label{D_mu(tildeB_mu)=0vsYs}
Y^s = -\frac{1}{2}S^\dagger (D_\mu B^\mu) S,
\ee
where
\be\label{Y}
Y\equiv S^\dagger  B^\mu D_\mu S,
\ee
and where $Y^s \equiv \frac{1}{2}(Y+Y^\dagger )$ denotes the Hermitian part of $Y$. But the r.h.s. of (\ref{D_mu(tildeB_mu)=0vsYs}) is a Hermitian matrix. Hence the equation $Z=0$, which is just
\be
Y= -\frac{1}{2}S^\dagger (D_\mu B^\mu) S,
\ee
is equivalent to the conjunction of (\ref{D_mu(tildeB_mu)=0vsYs}) {\it and} the vanishing of the {\it antihermitian} part of $Y$, namely
\be\label{Y^a=0}
S^\dagger  B^\mu D_\mu S - (D_\mu S^\dagger )B^\mu S =0.
\ee
In particular, $Z=0$ implies (\ref{D_mu(tildeB_mu)=0vsYs}), and thus implies (\ref{D_mu(tildeB_mu)=0})$_2$. This completes the proof.\\

Thus, coming back from the modified Dirac equation to the normal one leads necessarily to a normal equation for which the current is conserved. Note that Eq. (\ref{Z=0}) is a first-order linear system of 16 independent PDE's for the 16 independent unknowns of the similarity matrix $S(X)$. So that, starting from fields $\gamma ^\mu ,A$ ``selected at random," solving this system (which should be possible with suitable boundary conditions) allows us to go to {\it admissible} fields $\tilde{\gamma} ^\mu ,\tilde{A}$. The meaning of the additional condition (\ref{Y^a=0}) will be soon clarified by \hyperref[Theorem4]{Theorem 4}. For now, recall that, if we start from a system of fields $\gamma ^\mu ,A$ that already satisfies condition (\ref{D_mu(B_mu)=0}), then the ``modified" Dirac equation actually coincides with the normal one. Hence, \hyperref[Theorem3]{Theorem 3} allows us to immediately answer question ii) put at the \hyperref[similarity-psi]{beginning} of this subsection:\\

\paragraph{Corollary 2.}\label{Corollary2} {\it Let us start from a set of fields $\gamma^\mu,A,B^\mu \equiv A\gamma^\mu$ that does satisfy the condition (\ref{D_mu(B_mu)=0}) for current conservation. In order that $\tilde{\psi} $ still obey (\ref{Dirac-general}) each time that $\psi$ obeys it already, it is necessary and sufficient that 
\be\label{Z=0-DmuBmu=0}
Z\equiv B^\mu D_\mu S =0.
\ee
Moreover, condition (\ref{D_mu(B_mu)=0}) is then preserved by the similarity transformation.}\\

Note that Eq. (\ref{Z=0-DmuBmu=0}) is the particular case $D_\mu B^\mu=0$ in Eq. (\ref{Z=0}), and is thus also a first-order linear system of 16 independent PDE's for the 16 independent unknowns of the similarity matrix $S(X)$.\\

Finally, the answer to question iii) is given by\\

\paragraph{Theorem 4.}\label{Theorem4}{\it In order that $\tilde{\psi} $ still obey the modified equation (\ref{Dirac-general-modified}), each time that $\psi$ obeys it already, it is necessary and sufficient that Eq. (\ref{Y^a=0}) be satisfied.}\\

{\it Proof.} Let us write that $\tilde{\psi} $ obeys Eq. (\ref{Dirac-general-modified}), and multiply this by $S$:
\be\label{Dirac-general-modified-tilde}
0= S\left[\tilde{\gamma} ^\mu D_\mu\tilde{\psi}+im\tilde{\psi}+\frac{1}{2}\tilde{A}^{-1}(D_\mu \tilde{B}^\mu )\tilde{\psi}\right].\\
\ee
On the other hand, the fact $\psi $ obeys Eq. (\ref{Dirac-general-modified}) is equivalent to Eq. (\ref{similarity-Dirac-modified}), thus to 
\be
0  =  S\left[\tilde{\gamma} ^\mu D_\mu +im\right]\tilde{\psi} +\gamma ^\mu (D_\mu S) \tilde{\psi} +\frac{1}{2}A^{-1}(D_\mu B^\mu )S\tilde{\psi}.
\ee
Hence, $\tilde{\psi} $ obeys (\ref{Dirac-general-modified}) at each time that $\psi$ obeys it already, iff:

\be\label{similarity-Dirac-modified-2}
\forall \tilde{\psi} \quad \gamma ^\mu (D_\mu S) \tilde{\psi} +\frac{1}{2}A^{-1}(D_\mu B^\mu )S\tilde{\psi}=\frac{1}{2}S\tilde{A}^{-1}(D_\mu \tilde{B}^\mu )\tilde{\psi}.
\ee
Inserting the expression (\ref{similarity-A-Bmu}) of $\tilde{A}$ and $\tilde{B}^\mu $, and thus computing $D_\mu \tilde{B}^\mu $, yields
\be
S\tilde{A}^{-1}(D_\mu \tilde{B}^\mu )= A^{-1}(D_\mu B^\mu )S + \gamma ^\mu (D_\mu S) + A^{-1}(S^\dagger )^{-1}(D_\mu S^\dagger) B^\mu S.
\ee
Thus, the characteristic condition (\ref{similarity-Dirac-modified-2}) rewrites as
\be
\gamma ^\mu (D_\mu S) = A^{-1}(S^\dagger )^{-1}(D_\mu S^\dagger) B^\mu S,
\ee
or (remembering that $A\gamma ^\mu \equiv B^\mu $):
\be
S^\dagger B ^\mu (D_\mu S) = (D_\mu S^\dagger) B^\mu S,
\ee
which is precisely Eq. (\ref{Y^a=0}). Q.E.D.\\

Let us summarize. If a similarity transformation takes the modified Dirac equation (\ref{Dirac-general-modified}) to the normal one (\ref{Dirac-general}), then it leads to a normal equation with {\it admissible fields} $\gamma ^\mu ,A$, {\it i.e.,} ones for which, in addition to the anticommmutation relation (\ref{Clifford}), the condition (\ref{D_mu(B_mu)=0}) for current conservation is satisfied (\hyperref[Theorem3]{Theorem 3}). And if, starting with admissible fields $\gamma ^\mu ,A$, the normal Dirac equation is covariant under the transformation, then  necessarily the transformed fields are admissible also \hyperref[Corollary2]{(Corollary 2)}. This provides the justification for the restriction to admissible fields. Finally, the condition (\ref{Z=0}) of \hyperref[Theorem3]{Theorem 3}, that allows one to come back from the modified to the normal Dirac equation, turns out to be equivalent to the conjunction of the condition (\ref{D_mu(B_mu)=0}) for current conservation, and of condition (\ref{Y^a=0}). The latter is just the one ensuring the covariance of the modified Dirac equation (\hyperref[Theorem4]{Theorem 4}).

\section{Hermiticity of the Hamiltonian}\label{Hamiltonian}

\subsection{The Hamiltonian operator and its frame dependence}\label{Frame-dependence}

The general Dirac Hamiltonian $\mathrm{H}$ is obtained by multiplying the general Dirac equation (\ref{Dirac-general}) by $\gamma ^0$ on the left, using the anticommutation formula (\ref{Clifford}). This puts the Dirac equation into Schr\"odinger form:
\be \label{Schrodinger-general}
i \frac{\partial \psi }{\partial t}= \mathrm{H}\psi,\qquad (t\equiv x^0),
\ee
with 
\be \label{Hamilton-Dirac-general}
 \mathrm{H} \equiv  m\alpha  ^0 -i\alpha ^j D _j -i(D_0-\partial_0),
\ee
and where the $\alpha ^\mu $ 's are given by Eq. (\ref{alpha}). One sees from the latter equation that, for TRD, $\alpha^{\mu \rho }_\nu \equiv (\alpha^\mu)^\rho _{\ \,\nu}$ is {\it not} a general tensor, in contrast with $\gamma ^{\mu \rho }_\nu \equiv (\gamma ^\mu)^\rho _{\ \,\nu}$---and that, however, it does behave as a spacetime tensor for purely spatial transformation of coordinates: 
\be\label{purely-spatial-change}
x'^0=x^0,\ x'^j=f^j((x^k)).
\ee

\vspace{3mm}
This is natural. Indeed, the rewriting of any linear wave equation in the Schr\"odinger form (\ref{Schrodinger-general}), in which the (linear) Hamiltonian operator H has to contain no time derivative, is based on a splitting of spacetime into space and time. Thus, a priori, the operators H and H$'$ corresponding to different spacetime coordinate systems should be {\it different,} in general. This can be checked from the transformation of the Schr\"odinger equation (\ref{Schrodinger-general}). Consider first the case that the wave function is a scalar, or transforms as a scalar (which is relevant to DFW). Then Eq. (\ref{Schrodinger-general}) transforms simply as
\be\label{H'-psi-scalar}
\mathrm{H}'\psi' \equiv i\frac{\partial \psi' }{\partial t'} =i\frac{\partial \psi }{\partial x^\mu }\frac{\partial x^\mu }{\partial t'}=\frac{\partial t }{\partial t' }\mathrm{H}\psi + i\frac{\partial \psi }{\partial x^j }\frac{\partial x^j }{\partial t'}.
\ee
In order that the Hamiltonians $\mathrm{H}$ and $\mathrm{H}'$  be equivalent operators, it must be that: i) $\partial x^j /\partial t'=0$, ii) $\mathrm{H}\psi $ transforms as a scalar: $(\mathrm{H}'\psi')((x'^\mu) ) =(\mathrm{H}\psi)((x^\nu )) $, and iii) $t'=t$, that is, $x'^0=x^0$.
\footnote{\
Since we set $t\equiv x^0$ (and $t'\equiv x'^0$), the ``true" time is rather $T\equiv x^0/c$. Note that $x^0\equiv cT$ is invariant under a change $T'=aT$. In other words, time scale changes {\it are} (fortunately) allowed.
}
The three foregoing conditions impose that we restrict the allowed transformations to be purely spatial coordinate changes (\ref{purely-spatial-change}). Now, for DFW, $\mathrm{H}\psi $ as obtained from Eq. (\ref{Hamilton-Dirac-general}) is indeed a spatial scalar \cite{A38}. Next, consider the case that the wave function is a four-vector (which is relevant to TRD). We have then
\be\label{H'-psi-vector}
(\mathrm{H}'\psi')^\mu \equiv i\frac{\partial \psi'^\mu  }{\partial t'} =i\frac{\partial}{\partial t' }\left(\frac{\partial x'^\mu }{\partial x^\nu }\psi ^\nu\right) =\frac{\partial x'^\mu }{\partial x^\nu }\times \left(i\frac{\partial \psi ^\nu}{\partial x^\rho  }\frac{\partial x^\rho  }{\partial t'}\right),
\ee
the latter for linear coordinate changes. Thus, in order that the new Hamiltonian $\mathrm{H}'$ that appears in Eq. (\ref{H'-psi-vector}) be an equivalent operator to $\mathrm{H}$, the expression above should now transform as a four-vector. This needs that
\be
i\frac{\partial \psi ^\nu}{\partial x^\rho  }\frac{\partial x^\rho  }{\partial t'}=i\frac{\partial \psi ^\nu}{\partial t}\equiv (\mathrm{H}\psi)^\nu,
\ee
or
\be
\frac{\partial x^\rho  }{\partial t'}=\delta ^\rho _0.
\ee
That is, again one must restrict oneself to spatial changes (\ref{purely-spatial-change}). Since this restriction applies already to linear coordinate changes, it must be imposed to all coordinate changes. One verifies that, under all spatial coordinate changes (\ref{purely-spatial-change}), $\mathrm{H}'\psi'$, thus defined as the transformation of the l.h.s. of the Schr\"odinger equation (\ref{Schrodinger-general}), is indeed the four-vector transformation of $\mathrm{H}\psi$. One then easily checks that $\mathrm{H}\psi $, as defined instead from the explicit expression (\ref{Hamilton-Dirac-general}) of the Hamiltonian, does transform as a four-vector under spatial changes (\ref{purely-spatial-change}) for TRD. Thus, for DFW and TRD as well, in order that the Hamiltonians $\mathrm{H}$ and $\mathrm{H}'$ before and after a coordinate change be equivalent operators, the coordinate change must be a spatial change (\ref{purely-spatial-change})---then, both sides of the Schr\"odinger equation (\ref{Schrodinger-general}) behave as a scalar for DFW, and as a four-vector for TRD. This is consistent with the fact that, both $\psi $ and  $\mathrm{H}\psi$ being wave functions in the same Hilbert space, both $\psi $ and  $\mathrm{H}\psi$ must transform the same way under ``allowed" coordinate transformations.\\

Thus, the Hamiltonian operator associated with a given wave equation depends on the {\it reference frame} F which is considered---the latter being understood here as an equivalence class of local coordinate systems (charts) on the spacetime V, modulo the purely spatial transformations (\ref{purely-spatial-change}). 
\footnote{\ 
This notion of a reference frame is formalized in Ref. \cite{A45}, together with the notion of the associated space manifold M, which is time-independent. This formalization needs that one restricts oneself to an open domain U in V, such that there is at least one chart $\chi :\mathrm{U}\rightarrow {\sf R}^4$, thus that covers U. Since a chart is, from its definition, a diffeomorphism of its domain onto its range in the arithmetic space ${\sf R}^4\simeq {\sf R}\times {\sf R}^3$, this means that we assume a local $1 \times 3$ decomposition of spacetime---as is indeed commonly assumed in works on quantum theory in a curved space-time \cite{Birrell-Davies1982,Fulling1989}. This decomposition writes $X\mapsto (x^0,x)$ with $x\in\ $M: see Sect. 4 in Ref. \cite{A45}, point i). 
}
Since the time coordinate $x^0$  is fixed, this is a more restrictive definition of a reference frame than in relativistic gravity, where changes $x'^0=f((x^\mu ))$ are allowed \cite{Cattaneo,L&L}. However, one may trace back this restriction to that effected by mechanics itself: {\it e.g.} in an inertial frame in a flat spacetime, the inertial time (synchronized according to the Poincar\'e-Einstein procedure) is naturally distinguished; accordingly, the quantum Hamiltonian will generally change if one selects another time coordinate, see Eq. (\ref{H'-psi-scalar}). As shown in Ref. \cite{A45}, the data of a reference frame F determines a three-dimensional ``space" manifold M, which is {\it the set of the world lines of the observers bound to }F---{\it i.e.,} whose spatial coordinates $x^j$ do not depend on the time $x^0$, in any chart of the class F. The charts of the class F, also called charts adapted to F, provide an atlas of M: the coordinates of the running element of M in such a chart are just the constant spatial coordinates $x^j$ of the corresponding world line.\\

Equation (\ref{Hamilton-Dirac-general}) reveals an important property of the TRD-2 connection (\ref{Delta-connection-etherframe}) in the preferred frame E. Namely, using the definition (\ref{Delta-connection-etherframe}) to set $D_0=\partial_0$ in Eq. (\ref{Hamilton-Dirac-general}) shows that the Hamiltonian operator for TRD-2, at each time $t$, depends only on the choice of gamma matrices and the spatial geometry at time $t$.

\subsection{The scalar product}

Ideally, we would like to define in a natural way a Hilbert space scalar product for wave functions, and to find that ``the Hamiltonian H is Hermitian for this scalar product." We now know that this property, like H itself, will likely depend on the reference frame, so we select one, F, with the associated ``space" manifold M. Note that H, because it does not involve time derivatives, primarily operates on {\it spatial} wave functions $\psi =\psi (x)$ with $x \in \mathrm{M}$. However, in general, H and the metric $g_{\mu \nu }$ do depend on the time $t\equiv x^0$. As noted in Ref. \cite{A38}, the scalar product for the Dirac equation (\ref{Dirac-general}) has necessarily the following general form:
\be \label{Hermitian-general}
(\psi  \mid \varphi  ) \equiv \int_\mathrm{M} (\psi  (x). \varphi  (x)) \ \dd {\sf V}(x),
\ee
where $(u.v)$ is a Hermitian product defined for ``arrays $u$ and $v$ of four complex numbers" (which in fact are either complex four-vectors or quadruplets of complex scalars, depending on whether TRD or DFW theory is considered), and where $\dd{\sf V}(x)$ at each time  $t$ is an arbitrary volume element defined on M. The latter has the form (at time $t$)
\be\label{dV}
\dd {\sf V}(x)= \sigma (t,x)\ \sqrt{-g(t,x )}\ \dd^ 3{\bf x}, \qquad g\equiv \mathrm{det}(g_{\mu \nu })
\ee
where $\sigma (t,x)$  is any spatial scalar field [{\it i.e.,} scalar under transformations (\ref{purely-spatial-change})].  Note then using formula (\ref{purely-spatial-change}) that the integral defining the scalar product (\ref{Hermitian-general}) is invariant under all coordinate transformations of M.  Also note that the volume element  $\sigma (t,x)\ \sqrt{-g(t,x )}\ \dd^ 3{\bf x}$ is the most general possible on M. \\

For a flat spacetime, in any Cartesian coordinate system $(x^\mu)$, the following scalar product has been identified:
\be \label{Hermitian-flat}
(\psi \parallel   \varphi  ) \equiv \int_{\mathrm{space}} (\psi ({\bf x}):\varphi({\bf x}))   \ \dd^ 3{\bf x},
\ee 
where $(u:v)$ is the positive-definite product (\ref{B-product}) with the corresponding {\it constant} hermitizing matrix $B$. It is fully satisfying in that case, since the Hamiltonian is always Hermitian for that product \cite{A40}. Hence, the sought after scalar product (\ref{Hermitian-general}) must coincide with (\ref{Hermitian-flat}) in a flat region of U, when it is endowed with Cartesian coordinates. More generally, for any given event $X \in \mathrm{U}$, and for wave functions of a priori bounded variation, vanishing outside a small neighborhood of $X$, the metric and the matrix $B$ may be considered constant, so that one should be able to approximate the exact product (\ref{Hermitian-general}) by the product (\ref{Hermitian-flat}), rewritten in a covariant form. Therefore, we must take $(\psi  (x). \varphi  (x))\equiv (\psi  (x): \varphi  (x))$ in (\ref{Hermitian-general}), and we are left with the mere choice of the volume measure ${\sf V}$ on M, thus with the choice of the scalar $\sigma $. An obvious possible choice, indeed the standard choice, is $\sigma \equiv 1$. So that the general form of the possible scalar product may be written, both in intrinsic form and in a chart $\tilde{\chi }$ on the space M, as
\be \label{Hermitian-general-sqrt-g}
(\psi  \mid \varphi  ) \equiv \int_\mathrm{M} \ (\psi  : \varphi)\  \dd {\sf V}=\int_{\tilde{\chi }(\mathrm{M})} \psi^\dagger A\gamma ^0  \varphi \ \sigma \ \sqrt{-g}\ \dd^ 3{\bf x}.
\ee

\vspace{4mm}
For DFW theory, recall that the matrices $\gamma ^\mu $, in particular $\gamma ^0$, are defined by Eq. (\ref{flat-deformed}), in which the matrix $(a^\mu _{\ \,\alpha} )$ must be invertible and satisfy
\be\label{compatible-anticom}
a^\mu _{\ \,\alpha} \,a^\nu _{\ \,\beta } \,\eta ^{\alpha \beta }=g^{\mu \nu }.
\ee
If we make the standard choice $\sigma \equiv 1$, and take for ``flat" matrices the standard Dirac matrices $\gamma ^{\sharp \, \mu }$, so that we may choose $A=\gamma ^{\sharp \, 0 }$ as previously noted, then, in that particular case, the scalar product (\ref{Hermitian-general-sqrt-g}) is the one stated by Leclerc \cite{Leclerc2006}. If, in addition, we have $a^0 _{\ \,j}=0$, we get $a^0 _{\ \,0}=\sqrt{g^{00}}$ from (\ref{compatible-anticom}), hence
\be\label{a^0_j=0case}
\sqrt{-g}\,A\,\gamma ^0 = \sqrt{-g\,g^{00}}\, {\bf 1}_4.
\ee
Now assume that $g_{0j}=0 \ (j=1,2,3)$, as is in particular the case for a static metric in adapted coordinates. Then we can certainly get $a^0 _{\ \,j}=0$, and moreover we have $g^{00}=1/g_{00}$, whence
\be\label{g_0j=0case}
\sqrt{-g}\,A\,\gamma ^0 = \sqrt{h}\, {\bf 1}_4,
\ee
where $h\equiv \mathrm{det}(h_{jk})$ is the determinant of the metric $h_{jk}=-g_{jk}\ (j=1,2,3)$ on the space M, induced by the spacetime metric $g_{\mu \nu }$.  Thus, in the case that $g_{0j}=0 \ (j=1,2,3)$, and using furthermore the standard Dirac matrices, the definition (\ref{Hermitian-general-sqrt-g}) coincides with that found ``natural" in Ref. \cite{A38} [Eqs. (21)--(23) there] for a static metric in DFW theory.

\subsection{Conditions for hermiticity and isometric evolution}\label{Conditions for hermiticity}

In quantum mechanics, the Hamiltonian operator H$(t)$ is an operator-valued function of time $t$, that represents the energy observable at time $t$.  There are three axioms in quantum mechanics that can be straightforwardly adapted to a curved spacetime as follows:  \\
 
\paragraph{Axiom (A).}\label{AxiomA}  {\it The Hilbert space scalar product (\ref{Hermitian-general-sqrt-g}) of any time-independent wave functions $\psi $ and $\varphi $ defined on the space manifold $\mathrm{M}$ is time independent.}\\  

\paragraph{Axiom (B).}\label{AxiomB}  {\it  For each time $t$, the Hamiltonian  $\mathrm{H}$ is a Hermitian operator with respect to the scalar product (\ref{Hermitian-general-sqrt-g}).}\\

\paragraph{Axiom (C).}\label{AxiomC}  {\it  The solutions of the Dirac equation (\ref{Schrodinger-general}) have an isometric evolution with respect to the scalar product (\ref{Hermitian-general-sqrt-g}).}\\  

It will be shown in this subsection that Axioms (A) and (B) together uniquely determine the Hilbert space scalar product, up to an inconsequential constant.\\

Note that these axioms are somewhat weaker than the self-adjointness and unitary evolution required in quantum mechanics.  Axiom (B) is a prerequisite for the energy operator $\mathrm{H}(t)$ to be self-adjoint for each time $t$.  Axiom (C), which guarantees that the Hilbert space norm is preserved in time, is a prerequisite for ensuring that the solutions of the Dirac equation (\ref{Schrodinger-general}) have a unitary evolution with respect to the scalar product (\ref{Hermitian-general-sqrt-g}).\\   

It can be shown that any pair of these axioms implies the third.  In particular, we note that if Axiom (A) is valid, then Axioms (B) and (C) are equivalent, as they are in ordinary quantum mechanics.  This observation follows immediately from differentiating Eq. (\ref{Hermitian-general-sqrt-g}) with respect to time to obtain 
\be\label{d_0 (psi | phi)}
\partial_0(\psi\mid \varphi ) = \int_{\mathrm{M}}\psi ^\dagger \,\partial_0[\sigma \sqrt{-g}A\gamma ^0]\ \varphi \
\dd^3{\bf x}  + ((\partial_0\psi)\mid  \varphi)+(\psi\mid  (\partial_0\varphi))
\ee
and then substituting for $\psi $  and $\varphi $ either time-independent wave functions or time-dependent solutions of the Dirac equation (\ref{Schrodinger-general}).  In the former case, from Axiom (A) we have $\partial_0(\psi\mid \varphi ) =0$, and since $\psi $ and $ \varphi $ are time-independent we get 
\be\label{d_0 (psi | phi)-spatial}
0 = \int_{\mathrm{M}}\psi ^\dagger \,\partial_0[\sigma \sqrt{-g}A\gamma ^0]\ \varphi \
\dd^3{\bf x}  + 0.
\ee
Since  $\psi $ and $\varphi $  can be chosen  as  arbitrary smooth, 4-component complex wave functions with compact supports in $\mathrm{M}$, the vanishing of that integral implies that
\be\label{validAxiomA}
\partial_0[\sigma \sqrt{-g}A\gamma ^0] = 0.
\ee
Then, in the latter case, for two solutions  $\psi $ and $\varphi $ of the Dirac equation (\ref{Schrodinger-general}), we may substitute Eq. (\ref{validAxiomA}) and $\mathrm{H}=i\partial_0$ in Eq. (\ref{d_0 (psi | phi)}), obtaining
\be\label{d_0 (psi | phi)-solutions}
\partial_0(\psi\mid \varphi ) = i[(\mathrm{H}\psi\mid  \varphi)-(\psi\mid \mathrm{H} \varphi)].
\ee
Thus, assuming Axiom \hyperref[AxiomA]{(A)}, {\it the hermiticity of the Hamiltonian is equivalent to the isometricity of the evolution,} generalizing a well known result in quantum mechanics.\\

The hermiticity condition for the Hamiltonian is, by definition,
\be \label{Hermiticity-definition}
(\mathrm{H} \psi  \mid \varphi  ) =(\psi  \mid \mathrm{H} \varphi  )
\ee
for all $\psi $ and $\varphi $ in the domain of H$(t)$, denoted Dom(H), which we assume to be independent of time. Recall that the wave functions $\psi$ and $\varphi $ in this definition are time-independent. Because the sesquilinear form $(\mathrm{H} \psi  \mid \varphi  )$ is determined by the corresponding quadratic form $Q(\psi )\equiv ( \mathrm{H}\psi  \mid \psi   )$, an equivalent condition to (\ref{Hermiticity-definition}) is actually
\be \label{Hermiticity-definition-psi-only}
\forall \psi \in \mathrm{Dom(H)}\quad (\mathrm{H} \psi  \mid \psi   ) =(\psi  \mid \mathrm{H} \psi   ) \qquad (\mathrm{for\ each\ time\ } t).
\ee
In order to find when this does occur with the Hamiltonian (\ref{Hamilton-Dirac-general}), we use the expression of the global scalar product (\ref{Hermitian-general-sqrt-g}) as an integral of the Hermitian product $(u:v)$, and we use the relation between the two Hermitian products $(u:v)$ and $(u,v)$ in Eq. (\ref{:to,}), exploiting their Hermitian property to obtain:
\bea\label{delta-hermit}
(\psi : \mathrm{H}\psi )-(\mathrm{H}\psi :\psi )& = &(\psi :-i\alpha^j D_j\psi )+(i\alpha^j D_j\psi:\psi )\nonumber\\
& & +(\psi :-i(D_0-\partial_0)\psi )+(i( D_0-\partial_0)\psi:\psi )\nonumber\\
& = & (\psi ,-i\gamma ^j D_j\psi )+(i\gamma^j D_j\psi,\psi )+(\psi ,-i\gamma^0 D_0\psi )\nonumber\\
& & +(i\gamma^0 D_0\psi,\psi )+(\psi :i\partial_0\psi )-(i\partial_0\psi:\psi )\nonumber\\
&= & (\psi ,-i\gamma ^\mu  D_\mu \psi )+(i\gamma^\mu  D_\mu \psi,\psi)\nonumber\\
& & +(\psi :i\partial_0\psi )-(i\partial_0\psi:\psi ).
\eea
Therefore, if the wave function $\psi $ is a time-dependent one that obeys Dirac equation in the Schr\"odinger form (\ref{Schrodinger-general}), we get
\be
0= (\psi ,-i\gamma ^\mu  D_\mu \psi )+(i\gamma^\mu  D_\mu \psi,\psi ),
\ee
which is indeed an immediate consequence of the Dirac equation in the initial form (\ref{Dirac-general}). Thus we are left with $0=0$ in that case. \\

However, if the wave function $\psi $ is a {\it time-independent} one instead, as is normal at the stage of checking the hermiticity of the Hamiltonian, what we get is more interesting:
\bea \label{delta-hermit-time-independent}
(\psi : \mathrm{H}\psi )-(\mathrm{H}\psi :\psi )& = & (\psi ,-i\gamma ^\mu  D_\mu \psi )+(i D_\mu \psi,\gamma^\mu \psi )\nonumber\\
& = & -i[\psi ^\dagger A \gamma^\mu  D_\mu \psi +( D_\mu \psi ) ^\dagger A\gamma ^\mu \psi].
\eea
Now, recall that the field $\gamma ^\mu ,A$ may not be selected freely, but instead should satisfy the {\it admissibility condition} (\ref{D_mu(B_mu)=0}) [and relation (\ref{Clifford})]. If this is the case, we find thus [cf. Eq. (\ref{J-mu-standard-matrix})]:
\be \label{delta-Hpsi-admissible}
(\psi : \mathrm{H}\psi )-(\mathrm{H}\psi :\psi )= -iD_\mu (\psi ^\dagger A \gamma^\mu  \psi) \equiv -iD_\mu J^\mu.
\ee
As noted after Eq. (\ref{DmuJmu-Dirac}), for TRD, as well as for DFW, the connection acting on the probability current $J^\mu $ is the Levi-Civita connection, hence 
\be\label{D_mu(J^mu)}
\sqrt{-g} \ D_\mu J^\mu = \partial_\mu \left( \sqrt{-g}\ J^\mu \right).
\ee
We use the form (\ref{Hermitian-general-sqrt-g}) of the scalar product. With this scalar product, by integrating (\ref{delta-Hpsi-admissible}) over the space manifold, we get (setting the boundary term equal to zero by assuming that the functions $\psi \in \mathrm{D}$ decrease quickly enough  at spatial infinity):
\footnote{\ 
From now on, for definiteness, we shall assume that $\tilde{\chi }(\mathrm{M})={\sf R}^3$ (for the chart $\chi \in \ $F that we consider), and hence that the ``space" manifold M is diffeomorphic to ${\sf R}^3$. 
}
\bea 
i[(\psi \mid  \mathrm{H}\psi )-(\mathrm{H}\psi \mid \psi )] & = & \int_{{\sf R}^3} \sigma\ \sqrt{-g}\, D_\mu J^\mu \, \dd ^3{\bf x}\label{delta-hermit-admissible-init}\\
& = & \int_{{\sf R}^3} \sigma\ \partial _\mu \left(\sqrt{-g}\, J^\mu\right) \, \dd ^3{\bf x}\nonumber\\
& = & \int_{{\sf R}^3} \left[ \sigma\ \partial _0 \left(\sqrt{-g}\, J^0\right) -(\partial_j \sigma)\ \sqrt{-g}\, J^j \right] \dd ^3{\bf x} +\int_{{\sf R}^3} \partial _j \left(\sigma\,\sqrt{-g}\, J^j\right) \, \dd ^3{\bf x}\nonumber\\
& = & \int_{{\sf R}^3} \left[ \sigma\ \partial _0 \left(\sqrt{-g}\, \psi ^\dagger A \gamma^0   \psi\right)-(\partial_j \sigma)\ \sqrt{-g}\, \psi ^\dagger A \gamma^j  \psi \right] \dd ^3{\bf x} +0,\nonumber\\
i[(\psi \mid  \mathrm{H}\psi )-(\mathrm{H}\psi \mid \psi )]& = & \int_{{\sf R}^3} \psi ^\dagger \left[ \sigma\ \partial _0 \left(\sqrt{-g}\,  A \gamma^0  \right)-(\partial_j \sigma)\ \sqrt{-g}\, A \gamma^j  \right] \psi \, \dd ^3{\bf x}.\label{delta-hermit-admissible}
\eea
\noindent Using again the fact that a sesquilinear form is determined by the corresponding quadratic form, we find thus that the hermiticity of the Hamiltonian is equivalent to ask that, for all $\psi $ and $\varphi $ in $\mathrm{Dom}(\mathrm{H})$,
\be
\int_{{\sf R}^3} \psi ^\dagger N \varphi \, \dd ^3{\bf x}=0, \qquad N\equiv  \sigma\ \partial _0 \left(\sqrt{-g}\,  A \gamma^0  \right)-(\partial_j \sigma)\ \sqrt{-g}\, A \gamma^j.
\ee
In the same way as after Eq. (\ref{d_0 (psi | phi)-spatial}), the matrix $N$ must thus vanish for every ${\bf x} \in {\sf R}^3$. Hence,  the characteristic condition of hermiticity for the general form (\ref{Hermitian-general-sqrt-g}) of the scalar product is
\be\label{hermiticity-sigma}
 \sigma\ \partial _0 \left(\sqrt{-g}\,  A \gamma^0  \right)-(\partial_j \sigma)\ \sqrt{-g}\, A \gamma^j =0 \quad \mathrm{for\ every\ } {\bf x} \in {\sf R}^3.
\ee
This result opens the possibility that, in a given metric and with a given admissible set of fields $\gamma ^\mu , A$, one might get the Hamiltonian Hermitian by an appropriate choice of the scalar field $\sigma$, that determines the scalar product. However, combining Eq. (\ref{hermiticity-sigma}) with Eq. (\ref{validAxiomA}) shows that $\left(\sqrt{-g}\,  A \gamma^\mu   \right)\,\partial_\mu \sigma=0$.  That is, since the matrix $\sqrt{-g}\,  A $ is invertible, $\gamma^\mu   \,\partial_\mu \sigma=0$ . Then, since the gamma matrices are independent, $\partial_\mu \sigma =0$, proving that the scalar field $\sigma $  is indeed constant, assuming Axioms \hyperref[AxiomA]{(A)} and \hyperref[AxiomB]{(B)}.  Without loss of generality we may henceforth set $\sigma \equiv 1$. Thus we can state the \\

\paragraph{Theorem 5.}\label{Theorem5} {\it Axioms \hyperref[AxiomA]{(A)} and \hyperref[AxiomB]{(B)} uniquely fix the scalar product to be}
\be \label{Hermitian-sigma=1-g}
(\psi  \mid \varphi  ) \equiv \int_{{\sf R}^3} \psi^\dagger A\gamma ^0  \varphi \ \ \sqrt{-g}\ \dd^ 3{\bf x}.
\ee

\vspace{4mm}
Moreover, rewriting (\ref{hermiticity-sigma}) with $\sigma \equiv 1$ yields the following result: \\

\paragraph{Theorem 6.}\label{Theorem6} {\it Assume that the coefficient fields $\gamma ^\mu ,A$ satisfy the two admissibility conditions (\ref{Clifford}) and (\ref{D_mu(B_mu)=0}). In order that the Dirac Hamiltonian (\ref{Hamilton-Dirac-general}) be Hermitian (at time $t$) for the scalar product (\ref{Hermitian-sigma=1-g}), it is necessary and sufficient that
\be\label{hermiticity-condition-}
N({\bf x}) \equiv \partial _0 M({\bf x}) =0 \quad \mathrm{for\ every\ } {\bf x} \in {\sf R}^3,\qquad M \equiv  \sqrt{-g}\,  A \gamma^0.  \
\ee
}

The matrices in Eq. (\ref{hermiticity-condition-}) depend a priori on the chart, say $\chi $. However, as we saw, the Hamiltonian (\ref{Hamilton-Dirac-general}) and the scalar product (\ref{Hermitian-general-sqrt-g}) depend only on the reference frame F, {\it i.e.,} on the equivalence class to which the chart belongs, modulo a relation of the kind ``$\chi \mathcal{R} \chi '$ iff $\chi $ and $\chi'$ exchange by a purely spatial transformation (\ref{purely-spatial-change})" \cite{A45}. The transformation of the matrices in Eq. (\ref{hermiticity-condition-}) is consistent with this: it is easy to check that the validity or invalidity of the condition (\ref{hermiticity-condition-}) is invariant under any spatial coordinate change (\ref{purely-spatial-change}), thus it depends only on the reference frame F.\\

\subsection{Effect of an admissible change of the coefficient fields}\label{similarity-on-coefficients}

In a given spacetime $(\mathrm{V},g_{\mu \nu })$, there are infinitely many coefficient fields $(\gamma ^\mu ,A)$ that satisfy the two admissibility conditions (\ref{Clifford}) and (\ref{D_mu(B_mu)=0}). The question thus arises, whether or not (in a given reference frame F) the hermiticity condition (\ref{hermiticity-condition-}) is preserved by a change of the admissible coefficient fields---such changes will be called {\it admissible} changes of coefficient fields. We have the definite answer to this question only for DFW. For both DFW and TRD, we know that, if a change of the fields $\gamma ^\mu $ respects the anticommutation relation (\ref{Clifford}), then this is a similarity transformation (\ref{similarity-gamma-invert}), and that then the matrix $A$ and the field $B^\mu $ change according to (\ref{similarity-A-Bmu}). Therefore, the matrix $M$ of \hyperref[Theorem6]{Theorem 6} changes like $A$ and $B^\mu $:
\be \label{similarity-M}
\tilde{M}\equiv \sqrt{-g}\,  \tilde{A} \tilde{\gamma}^0 =S^\dagger  M S.
\ee
As an immediate consequence of this and \hyperref[Theorem6]{Theorem 6}, we have the\\

\paragraph{Corollary 3.} \label{Corollary3} {\it Let $(\gamma ^\mu,A) $ be a set of admissible coefficient fields. Let us fix the reference frame, so that the Hamiltonian operator and the scalar product are determined by the fields $(\gamma ^\mu,A) $. Let $S=S(x)$ be a space-dependent matrix such that condition (\ref{D_mu(B_mu)=0}) is still valid after the transformation (\ref{similarity-gamma-invert})--(\ref{similarity-A-Bmu}) of the coefficient fields. Then, the initial Hamiltonian is Hermitian iff the transformed Hamiltonian is Hermitian.}\\

In short: the hermiticity is preserved by admissible changes that do not depend on time. However, admissible changes may well depend on time.  For DFW, the admissible changes are restricted to local similarity transformations belonging to the spin group, $S(X) \in \sf{Spin(1,3)}$.
\footnote{\
Any such similarity is obtained from a change of the tetrad field $u_\alpha =a^\mu _{\ \, \alpha }\,\partial _\mu$, by a (proper) local Lorentz transformation $L=L(X)\in {\sf SO(1,3)}$:
\be
\tilde{u}_\beta  =L^\epsilon  _{\ \, \beta  }\,u _\epsilon   =L^\epsilon   _{\ \, \beta  }\,a^\mu _{\ \, \epsilon  }\,\partial _\mu =\tilde{a}^\mu _{\ \, \beta  }\,\partial _\mu,
\ee
thus
\be\label{atilde=a L}
\tilde{a}^\mu _{\ \, \beta  }=a^\mu _{\ \, \epsilon  }\,L^\epsilon   _{\ \, \beta  }.
\ee
This allows us to define gamma matrices from the same {\it fixed set } of ``flat" ones through these two tetrad fields [cf. Eq. (\ref{flat-deformed})]:
\be\label{flat-to-curved}
\gamma ^\mu =a^\mu _{\ \, \alpha }\,\gamma^{\natural \alpha }, \qquad \tilde{\gamma} ^\mu =\tilde{a}^\mu _{\ \, \beta  }\,\gamma^{\natural \beta }, \qquad A=\tilde{A}\equiv A^\natural.
\ee
Using (\ref{atilde=a L}) and (\ref{flat-to-curved}) together with the characteristic property of the spinor representation $L\mapsto \pm \mathsf{S}(L)$ (defined up to a sign), it is easy to show that we have 
\be\label{S=S(L)}
\tilde{\gamma} ^\mu =a^\mu _{\ \, \alpha }\,S^{-1}\gamma^{\natural \alpha }S=S^{-1}\gamma ^\mu S,\qquad S\equiv \pm \mathsf{S}(L).
\ee
Note that $L$ and $S\equiv \pm \mathsf{S}(L)$ can depend on the event $X$, thus in particular on the time $t$. Moreover, with the restriction of $S$ to the spinor representation: $S\equiv \pm \mathsf{S}(L)$ with $L \in {\sf SO(1,3)}$, one may show that $S^\dagger A^\natural\ S=A^\natural$, thus (\ref{flat-to-curved})$_3$ is compatible with (\ref{similarity-A-Bmu}).
}  
We know from \hyperref[Corollary1]{Corollary 1} that all such transformations, including time-dependent ones, are admissible for DFW, since they preserve the conditions (\ref{flat-deformed}), (\ref{D_mu(B_mu)=0}), as well as (\ref{Clifford}). Thus, the hermiticity condition is {\it always} (\ref{hermiticity-condition-}), for DFW (with the choice $A=A^\natural$). Therefore, the hermiticity conditions before and after an arbitrary similarity transformation $S$, are respectively
\be \label{hermiticity-starting}
\partial_0 M=0 \quad \mathrm{for\ every\ } {\bf x} \in {\sf R}^3
\ee
and \be \label{hermiticity-after-simil}
\partial_0(S^\dagger  M  S)=0 \quad \mathrm{for\ every\ } {\bf x} \in {\sf R}^3.
\ee
Obviously, if the initial matrix $M$ is independent of time, then, with a time dependent similarity transformation $S$, the transformed matrix  $\tilde{M}=S^\dagger  M  S$ will in general depend on time---thus contradicting $\partial_0 \tilde{M}=0$.\\

To take a sensible example, consider the very general case of an admissible coordinate system, {\it i.e.,} a coordinate system such that $g_{0 0}>0$ and such that matrix $(g_{j k})$ is negative definite. In that case, the lemma in Appendix B of Ref. \cite{A40} shows that the matrix $(a^\mu _{\ \,\alpha} )$ of Eq. (\ref{compatible-anticom}) (which is matrix $M$ in Ref. \cite{A40}) may be chosen to satisfy $a^0_{\ \,j}=0$, so that, taking for ``flat" matrices the standard Dirac matrices $\gamma ^{\sharp \, \mu }$, with $A=\gamma ^{\sharp \, 0 }$, we have [Eq. (\ref{a^0_j=0case})]
\be\label{M-a^0_j=0case}
M = \sqrt{-g\,g^{00}}\, {\bf 1}_4.
\ee
Thus, the hermiticity condition with that starting system, Eq.  (\ref{hermiticity-starting}), rewrites as Leclerc's condition \cite{Leclerc2006}:
\be\label{Leclerc-2}
\partial _0 (\sqrt{-g \,g^{00}})=0\quad \mathrm{for\ every\ } {\bf x} \in {\sf R}^3,
\ee
while in the transformed system, it is (\ref{hermiticity-after-simil}), or here:
\be \label{hermiticity-similarity}
\partial_0(\sqrt{-g\,g^{00}} \,S^\dagger  S)=0 \quad \mathrm{for\ every\ } {\bf x} \in {\sf R}^3.
\ee
Now, if (\ref{Leclerc-2}) is satisfied, and if $S$ is such that $S^\dagger S$ does depend on time, it follows that (\ref{hermiticity-similarity}) {\it cannot} be satisfied. This proves that, {\it for DFW, the hermiticity condition is not invariant by the admissible changes of the coefficient fields $(\gamma ^\mu ,A)$.} \\

For TRD, the condition (\ref{D_mu(B_mu)=0}) is much more demanding than it is for DFW, because: i) The derivatives $D_\mu \gamma ^\nu $ have no reason to vanish in general, and ii) also the hermitizing matrix $A$ is not covariantly constant in general. For instance, in the case of a flat spacetime, the condition $D_\mu \gamma ^\nu =0$ means (for TRD) that the gamma matrices are constant in Cartesian coordinates. In contrast, for DFW, the condition $D_\mu \gamma ^\nu =0$ is always satisfied, so that in Cartesian coordinates the gamma matrices need not be constant. 

\section{Summary and conclusion}\label{Conclusion}

In this work, we studied simultaneously three versions of the Dirac equation in a curved spacetime: the standard or ``Dirac-Fock-Weyl" (DFW) equation, and two tentative versions, proposed recently \cite{A39}. In a given coordinate system, the three equations differ merely in the covariant derivative. The two alternative versions are based on the {\it tensor representation} of the Dirac (TRD) field, Eqs. (\ref{psi-vector}) and (\ref{gamma-(^2_1)tensor}). In order to define a conserved probability current for any one of these Dirac equations in a curved spacetime, we must consider a general set of Dirac gamma matrices $(\gamma ^\mu )$ and a hermitizing matrix $A$.  We call these matrices  $(\gamma ^\mu ,A)$, which vary with each spacetime point, {\it the coefficient fields of the Dirac equation}.  Different choices for the coefficient fields $(\gamma ^\mu ,A)$ are related by local similarity transformations. For the two alternative equations based on TRD, these local similarity transformations are not restricted to $\mathsf {Spin(1,3)}$ transformations (associated with Lorentz transformations of a tetrad \cite{BrillWheeler1957+Corr}), nor to the unitary transformations which are considered in Ref. \cite{Pal2007}, but instead comprise the entire group of $\mathsf {GL(4,C)}$ transformations. \\

Independently of which Dirac equation is selected, the current conservation asks for an {\it admissibility condition} to be satisfied by the coefficient fields $(\gamma ^\mu,A)$, as shown in \hyperref[Theorem1]{Theorem 1}. This condition restricts only the choice of $A$ for DFW and is thus essentially always verified for DFW \hyperref[Corollary1]{(Corollary 1)}. But it also strongly restricts the choice of the field $\gamma ^\mu $ for TRD. However, one may {\it modify} the gravitational Dirac equations (any of them) so that the current conservation is always satisfied, as proved in \hyperref[Theorem2]{Theorem 2}. Moreover, starting from any coefficient fields $(\gamma ^\mu,A)$, one may transform to $(\tilde{\gamma} ^\mu,\tilde{A})$ satisfying the admissibility condition, by a local similarity transformation---this is shown by \hyperref[Theorem3]{Theorem 3}. For TRD, the local similarity transformations under which the ``unmodified" or the ``modified" Dirac equation is covariant are characterized by \hyperref[Corollary2]{Corollary 2} and \hyperref[Theorem4]{Theorem 4}, respectively.\\

We prove in \hyperref[Theorem5]{Theorem 5} that the Hilbert space scalar product is fixed by the axioms of quantum mechanics. Note in passing, that some authors choose a Hilbert space scalar product that differs from Eq. (\ref{Hermitian-sigma=1-g}), and therefore their Hilbert space does not satisfy the axioms of quantum mechanics listed as Axioms \hyperref[AxiomA]{(A)}, \hyperref[AxiomB]{(B)}, and \hyperref[AxiomC]{(C)} in Section \ref{Conditions for hermiticity} \cite{Audretsch1974}. The hermiticity of the Hamiltonian for the scalar product compatible with the axioms is studied in the general case in \hyperref[Theorem6]{Theorem 6}, {\it i.e.,} for a general coordinate system in a general curved spacetime. In particular, in any time-independent metric, the Hamiltonian is Hermitian for any admissible choice of time-independent coefficient fields $(\gamma ^\mu,A)$. However, for the standard equation (DFW), the Hamiltonian in a given coordinate system may be Hermitian for some admissible choice of the coefficient fields $(\gamma ^\mu,A)$, and non-Hermitian for another admissible choice---see Subsection \ref{similarity-on-coefficients}. It means that the very existence of an isometric evolution of the states (Axiom \hyperref[AxiomC]{(C)}) depends on an arbitrary choice of coefficient fields  $(\gamma ^\mu,A)$---for DFW.  \\

One may ask if the use of the TRD version of the Dirac equation leads to a clearer and more transparent presentation of the standard DFW formalism.  The answer is yes.  Note that the quantum mechanical treatment of the Dirac equation today is mostly limited to stationary spacetimes \cite{Fulling1989}.  The reason for this, as discussed in Section \ref{Hamiltonian}, is that in a non-stationary spacetime, the time evolution of free particle wave functions is generally not unitary, since the Dirac Hamiltonian is not generally Hermitian.  By investigating a wider group of transformations of the Dirac equation, the possibility exists that one can extend quantum mechanics to non-stationary spacetimes, so that the time evolution of free particle wave functions is unitary.  To explore this the wider group of local similarity transformations is a key ingredient. \\



\end{document}